# Induced superconductivity in magic-angle twisted trilayer graphene through graphene-metal contacts

Shujin Li[†*], Guanyuan Zheng[†*], Junlin Huang[†*]

Vanke Meisha Academy, Shenzhen, Guangdong 518083, China.

[†] These authors contributed equally
[*] Corresponding authors: lishujin@stu.vma.edu.cn (S.L.), liguanyuan@stu.vma.edu.cn (G.Z.), huangjunlin@stu.vma.edu.cn (J.H.)

**Abstract**

Magic-angle twisted trilayer graphene (MATTG) recently exhibited robust superconductivity at a higher transition temperature ($T_C$) than the bilayer version. With electric gating from both the top and bottom sides, the superconductivity was found to be closely associated to two conditions: the finite broken mirror symmetry and carrier concentrations between two to three carriers per moiré unite cell. Both conditions may be achieved by graphene-metal contacts where charge transfers and interfacial electric fields are generated to balance work function mismatch. In this study, we explore the superconductivity of MATTG when contacting a metal, through self-consistently solving the interfacial charge transfer with a highly electric-field-dependent band structure of MATTG. The predicted $T_C$ of MATTG-metal contacts forms two domes as a function of the work function difference over the interface, with a maximum over 2 K. Our work provides a constructive reference for graphene experiments and industrial applications with graphene-metal and graphene-semiconductor contacts.

**Introduction**

The recent advance of experimental control of moiré superlattices through stacking two or more layers of two-dimensional materials with twisted angles or lattice mismatch has attracted great attentions[1–3]. Such moiré superlattice forms the so called flat bands, leading to the observation of rich correlated and topological electronic phases[4–24]. Based on the pioneering discovery of superconductivity in the twisted graphene bilayers[4,5,17], the design of a trilayer graphene structure, where the carrier concentration and the mirror symmetry breaking are both at play, has led to more robust and higher superconducting transition temperature ($T_C$) (ref.[13,14,25]). The trilayer graphene, as a system with weak spin-orbit coupling[26], exhibits superconductivity after twisting and biasing with in-plane magnetic field far exceeding the Pauli limit[25], indicating an unconventional origin of the Cooper pairing.

To achieve the superconductivity, the magic-angle twisted trilayer graphene (MATTG) is stacked with the middle layer twisted for $\theta \sim 1.6$ degrees[13,14]. This angle is larger than the magic angle for bilayers, because the MATTG flat bands can be mathematically reduced to be similar to the bilayer version with effective twist angle of $\theta/\sqrt{2}$ at zero mirror symmetry breaking[27–30]. Dual electric gating from both sides of the trilayer enables an individual tunability of both the carrier concentration and the mirror symmetry breaking by altering the displacement field ($D$) across the trilayer graphene. At finite $D$, the flat bands hybridize with the Dirac bands, resulting in a change of bandwidth and interaction strength in the flat bands. The highest $T_C$ of MATTG can be achieved to exceed 2K on the hole doped side, as compared to about 1K of its bilayer version.

Since the electric displacement field and carrier density are two vital aspects for the superconductivity of MATTG, another way to naturally generate such electric displacement field together with carrier transfer is by contacting the MATTG with a metal[31–36]. When a metal contacts the MATTG, their difference in work function causes electrons to move across the two materials until the chemical potential got balanced when the system reaches the lowest possible energy. The charge transfer will then cause an interfacial electric field at the contact. Graphene-metal contacts widely exist in graphene related quantum devices and experiments. However, the impact on the superconductivity of MATTG-metal contact has been underexplored. In this study, we investigate the superconducting $T_C$ of MATTG when contacting metals and their alloys by self-consistent solving the charge transfer and interfacial electric displacement field considering the highly field-dependent flat band structure.

**Calculation Methods**

Our theoretical model consider the shift of Fermi level $\Delta E_F$ of the MATTG when it contacts a clean and flat metal surface. When the work function of the metal is different from the MATTG, as illustrated in Fig.1, a charge transfer between the metal and the graphene occurs. The Fermi level of the MATTG will be shifted, and simultaneously, an interfacial electric field is built up due to the formation of the interfacial dipole layers. Figure 1 demonstrates a case in which the work function of the MATTG is low than the metal. To balance the chemical potential, electrons form the MATTG flows to the surface of the metal, resulting a hole doping in the MATTG. The transfer of charges forms a dipole layer at the interface, and further establishes an electric field. When the metal has a lower work function than the MATTG, the case is reversed, where charges transfer occurs in the opposite direction and thus the electric field also points to a opposite direction. The relation between the carrier density and the Fermi level shift in the MATTG can be described as

$$n = 4 \cdot n_s \int_0^{\Delta E_F} \mathrm{DOS}(E, D)\, dE, \qquad (1)$$

where $n_s = \frac{2\theta^2}{\sqrt{3}a^2} \approx 1.5 \times 10^{16}\, m^{-2}$ corresponds to the density with one electron per moiré unit cell, in which $a = 0.246\, nm$ is the graphene lattice constant, and DOS(*E, D*) is the density of states (DOS) as a function of binding energy *E* and displacement field within the MATTG *D*.

We extract the density of state calculation results from reference[30]. The average displacement field *D* within the MATTG, based on the Gauss's law, can be estimated as

$$D = n \cdot e/2, \qquad (2)$$

where *e* is the charge of an electron. The factor 1/2 accounts for the reduced electric field within the MATTG due to the screening from the first and second graphene sheets. For instance, in a simplified estimation assuming that the charges are evenly distributed in the three layers of graphene, the displacement field in the two gaps between the three layers would be 2*ne*/3 and *ne*/3, giving an average of *ne*/2. The displacement field can be related to electric field with the relation $E = \frac{D}{\varepsilon_0}$.

We self-consistently solve equations (1) and (2) numerically to obtain relations $n(\Delta E_F)$ and $D(\Delta E_F)$. Since $T_C$ is a function of both $n$ and $D$, by extracting $T_C$ values from reference[14], we can therefore predict $T_C$ as a function of $\Delta E_F$.

**Results and Discussions**

Figure 2(a) shows the DOS of MATTG as a function of displacement field strength $D$ and binding energy $E$. The DOS is peaked at around $E = 3$ meV, where the flat bands are located, and rapidly decreases on both sides, where the Dirac bands are located. Apparently, the flat bands dominate the density of states and account for the majority of states near the $\Delta E_F$. With the increase of mirror symmetry breaking, represented by $D$, the flat bands hybridize with the Dirac bands, giving rise to: (i) a decrease of the peak value and (ii) an increase of bandwidth of the flat bands[30]. This could also induce stronger correlations between electrons which accounts for the emergent superconductivity[13,14,25].

Contacting MATTG with a metal surface induces charge transfer and, simultaneously, the mirror symmetry breaking. The normalized carrier density, $\nu = n/n_s$, and the displacement field $D$ in this metal-MATTG setup are not individually tuned, and the band structure is also altered instantaneously. With the shift of $E_F$, the variations of both variables are shown in Fig. 2(b). With the increase of $\Delta E_F$ to positive, $\nu$ increases indicating more electrons are filled and $D$ becomes negative, i.e., the electric field is pointing from the metal to the MATTG. On the negative $\Delta E_F$ side, $\nu$ decreases indicating holes and $D$ is pointing to the opposite direction. However, as can be clearly seen in Fig. 2(b), the positive and negative sides are not symmetric: on the positive $\Delta E_F$ side, both variables change more rapidly. This is because the peak of the DOS is on the positive side rather than at exact zero, as mentioned above.

Figure 3(a) and (b) shows the $T_C$ of MATTG as a function of $D$ and $\nu$, as extracted from reference[14]. Superconductivity can be observed when $|\nu|$ is between 2 and 3, and $T_C$ can be maximized with finite $D$. It can be also seen that the $T_C$ on the hole side is significantly higher than the electron side. With the $D$ and $\nu$ found in Fig. 2(b), we plot the line $D(\nu)$ across Fig. 3(a) and (b) to find the accessible phase space of MATTG-metal contact in the MATTG superconducting phase diagram. Interestingly, the $D(\nu)$ line almost crosses the peak of $T_C$ on the hole side. This means that contacting a metal would possibly cause the MATTG to superconduct at a temperature close to its maximum capability. By corresponding each $T_C$ value crossed by the $D(\nu)$ line with the $\Delta E_F$ value, we construct the $T_C(\Delta E_F)$ relation, as shown in Fig. 3(c). We see two superconducting domes in the investigated range. The one on the hole side has a highest $T_C$ exceeding 2 K, corresponding to $\Delta E_F = -10$ meV. The hole superconducting dome ranges from $\Delta E_F = -8$ meV to $-20$ meV. On the electron side, the dome is much smaller, ranging from $\Delta E_F = 5$ meV to 8 meV, with peak $T_C$ value of about 0.5 K.

Based on what we have found so far, can we predict what kind of metal would induce superconductivity for MATTG? In a simplified model, considering the DOS in metal is significantly higher than that of graphene, the graphene in contact with metal would shift its chemical potential to match the metals[37]. Under such an assumption, $\Delta E_F$ can be simply treated as equivalent to the work function difference between the MATTG and the metal. In this case, to induce superconductivity in MATTG on the hole side, we would want the metal to have very similar but about 10 meV higher work function than the MATTG (work function about 4.6 eV) (ref.[38]). To realize this metal, alloys between two metals with higher and lower work function than graphene can be designed, such as the alloy of Ag (work function about 4.3eV) and Au (work function about 5.2 eV). However, the actual charge transfer and Fermi level shift might dependent on much more complex conditions such as the distance between MATTG and the metal, the metal's surface crystalline orientation, and possibly the hybridization of bands between graphene and the metal[39]. Therefore, while our results provide a prediction of metal-induced superconductivity, the experimental realization would still require considerable trials. Our results are also useful for MATTG-semiconductor contacts. While charge transfer could result in either Schottky barrier or carrier accumulation at the surface of semiconductor, Fermi level shift together with interface dipole layer formation does not change and therefore our prediction still largely holds.

If metals are contacted on both sides of the MATTG, the situation would become more complex and could enable individual tunability of the carrier density and displacement field, since charge transfer not only occurs between the metals and graphene, but there could also be charge transfer between the two metals. In a simplified model where DOS of metals are assumed to be very large, to balance the chemical potential of the two metals, a dipole layer will form on the two surfaces of the metal, sandwiching the MATTG. The electric field between the two metal surfaces is proportional to the chemical potential difference of the two metals, and therefore can be tuned independently to some extent, but controlling the work function difference of these two metals. On the other hand, to match the balanced chemical potential of the two metals, the MATTG in between would also need to change its Fermi level. In this case, the carrier density of MATTG can be tuned roughly by the average work function of the two metals.

In conclusion, by self consistently solving the carrier density and displacement field in the magic angle twisted trilayer graphene given a Fermi level shift, we demonstrated that metal-MATTG contact could induce superconductivity, considering the hybridization and band width change with mirror symmetry breaking. The transferred charge between the metal and the MATTG produces an electric field on the MATTG that enables it to exhibit superconductivity features. As a function of Fermi level shift, we found two superconducting domes on both the hole and electron sides, with the highest $T_C$ surpassing 2 K. Our calculation predicts that a metal with similar but slightly (about 10 meV) higher work function than graphene would induce the highest $T_C$. Our results not only provides useful reference for future quantum devices incorporating MATTG, but also envision innovative ways for $T_C$ modulation.

**Acknowledgement**

We are grateful for the useful discussion with Ivan Tsang.

**Conflict of interest**

The authors declare no conflict of interests

**Author contributions**

S.L. performed theoretical analysis and wrote the manuscript. G.Z. and J.H. wrote the code and performed numerical calculations.

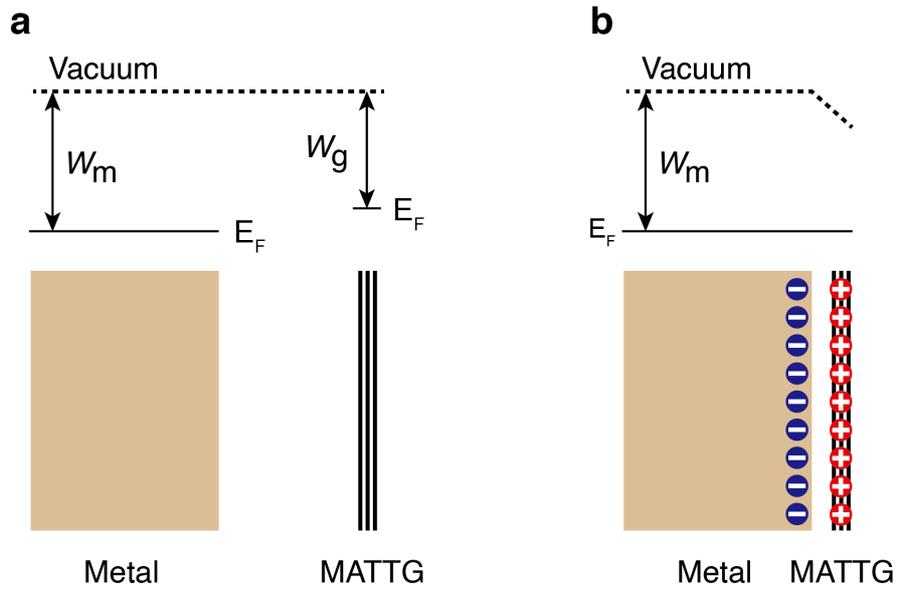

**Figure 1.** Schematic diagram of band alignment between a metal and the MATTG before contact (a) and after contact (b). $W_m$ and $W_g$ are the work functions of the metal and the MATTG, respectively. $E_F$ is the chemical potential. Dashed lines represent the vacuum energy level. Blue and red circles represent negative and positive charges, respectively.

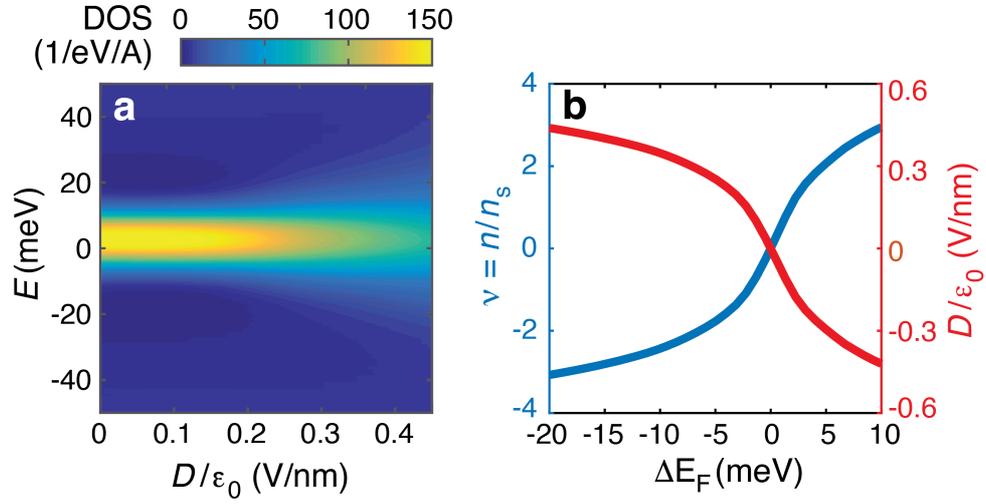

**Figure 2.** (a) Density of states as a function of electric displacement field within the MATTG and binding energy. A is the area of a moiré unit cell. (b) Carrier density (positive corresponds to electron-type) in the MATTG from interfacial charge transfer as a function Fermil level shift $\Delta E_F$ of MATTG due to the MATTG-metal contact (left axis, blue curve). Displacement field within the MATTG as a function of $\Delta E_F$ (right axis, red curve).

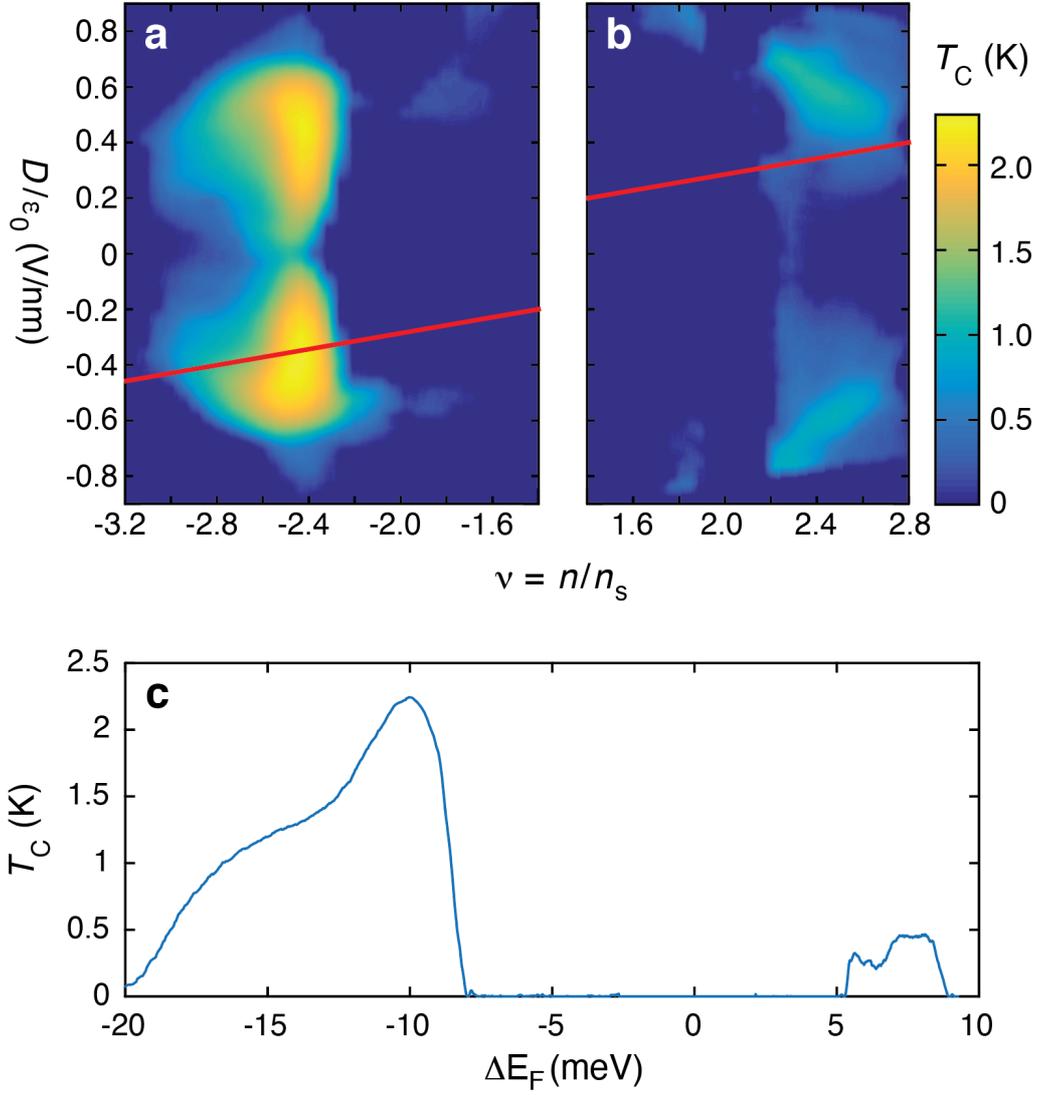

**Figure 3.** (a,b) $T_C$ as a function of displacement field and carrier density on the hole (negative, a) and electron (positive, b) sides. The red lines correspond to the displacement field and carrier density combinations accessible by MATTG-metal contact. (c) Extracted $T_C$ as a function of $\Delta E_F$ of MATTG due to the MATTG-metal contact.

17